\documentclass{aa}

\usepackage{graphicx}
\usepackage{txfonts}

\def\jcd{Christensen-Dalsgaard}
\newlength{\figwidth}
\begin{document}

\title{Temporal Variations in the Sun's Rotational Kinetic Energy}

\author{H. M.~Antia \inst{1}, S. M.~Chitre \inst{2} \and D. O.~Gough
\inst{3}}
\offprints{H. M.~Antia, \email{antia@tifr.res.in}}
\titlerunning{Temporal Variations in the Sun's Rotational Kinetic Energy}
\authorrunning{Antia, Chitre \& Gough}

\institute{Tata Institute of Fundamental Research, Homi Bhabha Road,
   Mumbai 400005, India \and
Department of Physics, University of Mumbai, Mumbai 400098, India, and
Institute of Astronomy, University of Cambridge, Madingley Road, Cambridge 
CB3 0HA
\and
Institute of Astronomy, University of Cambridge, Madingley Road, Cambridge 
CB3 0HA, and Department of Applied Mathematics and Theoretical Physics, 
Centre for Mathematical Sciences, Wilberfore Road, Cambridge CB3 0WA, UK}

\date{Received ??, 2005; accepted ??}

\abstract{}{
To study the variation of the angular momentum and the rotational 
kinetic energy of the Sun, and associated variations in the gravitational 
multipole moments, on a timescale of the solar cycle.}
{Inverting helioseismic rotational splitting data obtained by the Global 
Oscillation Network Group and by the Michelson Doppler Imager on the Solar 
and Heliospheric Observatory.}
{The temporal variation in angular momentum and kinetic energy at 
high latitudes ($>\pi/4   $) through the convection zone is positively 
correlated with solar activity, whereas at low latitudes it is anticorrelated, 
except for the top 10\% by radius where both are correlated positively.}
{The helioseismic data imply significant temporal variation in the 
angular momentum and the rotational kinetic energy, and in the gravitational 
multipole moments.  The properties of that variation will help constrain 
dynamical theories of the solar cycle.}

\maketitle

\section{Introduction}

Helioseismic data from the Global Oscillation Network Group (GONG)
and the Michelson Doppler Imager (MDI) have been used to infer the
rotation rate in
the solar interior (Thompson et al.~1996; Schou et al.~1998).
With the accumulation of these data for about 11 years it has now become possible
to study its temporal variations.  
Previous 
studies have 
revealed 
a distinct pattern of bands of faster and
slower than average rotation (Howe et al.~2000a; Antia \&
Basu 2000), which at low latitudes move towards the equator,
and at high latitudes move towards the poles (Antia \& Basu 2001). 
This pattern, which 
is similar
to the torsional oscillations observed at the solar surface
(Howard \& LaBonte 1980), 
has been well studied, and it has been found to penetrate through much of the body of the convection zone (Vorontsov et al.~2002;
Basu \& Antia 2003; Howe et al.~2005, 2006). Below the
convection zone, however, there is no generally accepted unambiguous detection
of significant temporal variation in the rotation, although
Howe et al.~(2000b, 2007) have reported a periodic variation near the equator 
between $r=0.65R_{\odot}$ and $r=0.75R_{\odot}$  in the early years of the 
last sunspot cycle with a period of 1.3 years, and Gough (2007) has reported
that there is evidence that the oscillation penetrates even more deeply. Other 
studies (Antia \& Basu 2000; 
Basu \& Antia 2003) have not confirmed such a variation.

A direct consequence
of temporal variation in the angular velocity is that global quantities
like angular momentum
and kinetic energy of rotation must also exhibit variation with time
(cf., Komm et al.~2003).
Because of the steep increase in density with depth, the major contribution
to these global quantities is naturally likely to arise from the deeper layers,
where a temporal variation in rotation has not been determined reliably;
and as a result, it is difficult to draw any firm 
conclusion 
about
the global variation of angular momentum and kinetic energy. However, following Komm et al.~(2003), we may consider
the contributions from different layers within the convection zone
separately.  We hope that such a study will shed some light on the
rotational dynamics and its role in driving the activity cycle.
Indeed, the study we present here is very similar to that of Komm et
al.~(2003), although we are now able to extend it over an entire solar cycle.
Apart from angular momentum and kinetic energy, we study also the temporal variation
of the gravitational quadrupole and higher-order multipole moments
due to the distortion in the solar figure induced by the known rotation.

In this work we concentrate principally on temporal variations in angular momentum
and rotational kinetic energy in different regions in the solar convection
zone, and compare their variation with the solar activity.  We describe
in Section 2 the basic technique and the data used in the study; the main
results are described in Section 3. In Section 4 we present the
variations in the multipole moments of the external gravitational
potential.  We draw our conclusions in Section 5.

\section{The data and their analysis}

We use the data from GONG (Hill et al.~1996) and SOI/MDI (Schou 1999). Each data set
consists of mean frequencies of different $(n,l)$ multiplets,
and the corresponding splitting coefficients. We use 110 temporally overlapping
data sets from GONG, each covering a period of 108 days, starting
from 1995 May 7 and ending on 2006 May 20, 
each set being displaced by 36 days from its predecessor. 
The MDI data consist
of 49 contiguous data sets, each covering a period of 72 days,
starting from 1996 May 1 and ending on 2006 May 20. These data
cover the rising phase of the cycle 23 and most of the descending
phase as well. We use a 2D Regularized Least-Squares (RLS) inversion
technique in the manner adopted by Antia et al.~(1998) to infer the
angular velocity in the
solar interior from each of the available data sets.

To study the temporal variation in the angular velocity we look at the
residuals obtained by subtracting from the angular velocity at any given
time its temporal mean $\Omega_0(r, \theta)$
\begin{equation}
\delta\Omega(r,\theta,t)=\Omega(r,\theta,t)- \Omega_0(r,\theta) \; ,
\end{equation}
with respect to the spherical polar coordinates $(r, \theta)$, where
$t$ is time.  The averaging to obtain $\Omega_0$ is over the entire time 
interval in which the data are
available.  Both the temporal mean and the residuals were calculated
separately for the GONG and MDI data.
The residuals, which essentially define the temporally varying
component of the angular velocity, were used (together with the means) 
to calculate the variations with epoch
in the kinetic energy and angular momentum in different regions
of the Sun. For example, the change in angular momentum of the
Sun between spherical surfaces at radii $r_1$ and $r_2$ can
be written as
\begin{equation}
\delta J(t; r_1,r_2)=2\pi\int_{r_1}^{r_2}\rho r^4\;{\rm d}r
\int_0^\pi \delta\Omega(r,\theta,t)\sin^3\theta\;{\rm d}\theta\;,
\end{equation}
where $\rho(r)$ is the density in the Sun. The integrand in the outer
integral defines the contribution to the angular momentum per unit radius
from an infinitesimal spherical shell around radius $r$, namely 
${\partial}\delta J/{\partial} r := {\partial}\delta J/{\partial} r_2|_{r_2=r}$. 
Similarly, we can also calculate the temporal variation in the
kinetic energy of rotation from the expression
\begin{eqnarray}
\lefteqn{\delta T(t; r_1, r_2) \simeq}\nonumber\\
& &\displaystyle 2\pi\int_{r_1}^{r_2}\rho r^4\;{\rm d}r
\int_0^\pi \Omega_0(r,\theta)\,\delta\Omega(r,\theta,t)\sin^3\theta\;{\rm d}\theta.
\end{eqnarray}
In evaluating expressions (2) and (3) we have used the density $\rho (r)$ 
from a standard solar model.

In addition to integrating over the entire latitude range  we can
consider also contributions from different latitude intervals. The observed
variation in angular velocity exhibits different behaviour in the
low- and high-latitude regions. At low latitudes the bands of more
rapidly and more slowly rotating fluid migrate towards the equator, while at high latitudes the bands move polewards
(Antia \& Basu 2001; Vorontsov et al.~2002; Howe et al.~2005;
Basu \& Antia 2006).
We therefore consider the contribution to the
integral separately
over the low latitudes ($\le\pi/4$) and high latitudes ($\ge\pi/4$)
to see how they compare with the corresponding global quantities.

The integrands in equations (2) and (3) are essentially the same, save
for the factor $\Omega_0$ in equation (3).  Therefore because the
angular variation of $\Omega_0$ is substantial only in the polar
regions where the factor ${\rm sin}^3\theta$ is quite small, the
temporal variations $\delta J$ and $\delta T$ are very similar.
Consequently in most of the following discussion we consider explicitly 
only the temporal variation in the kinetic energy.  We compare it with 
a solar activity index, for which we use the radio flux at 10.7 cm as a proxy.

Apart from kinetic energy and angular momentum, it is possible
to compute the gravitational quadrupole and higher-order multipole moments
of the Sun resulting from the centrifugal force (e.g. Schwarzschild 1947; 
Sweet 1950;  Gough 1981, 1982; Ulrich \& Hawkins 1981; Pijpers 1998; 
Antia et al.~2000; Roxburgh 2001;  Mecheri et al.~2004).  The gravitational 
potential $\Phi(r,\theta)$ outside the Sun can be written as
\begin{equation}
\Phi(r,\theta,t)=-{GM_\odot\over r}\left[1-\sum_{k=1}^\infty
\left(R_\odot\over r\right)^{2k} J_{2k}(t)P_{2k}(\cos\theta)\right],
\end{equation}
where $J_{2k}$ are the dimensionless multipole moments and
$P_{2k}$ are Legendre polynomials of degree $2k$.  We
attempt to study the temporal variations in these quantities.

\section{Results}

\begin{figure}[t]
\centerline{\resizebox{\figwidth}{!}{\includegraphics{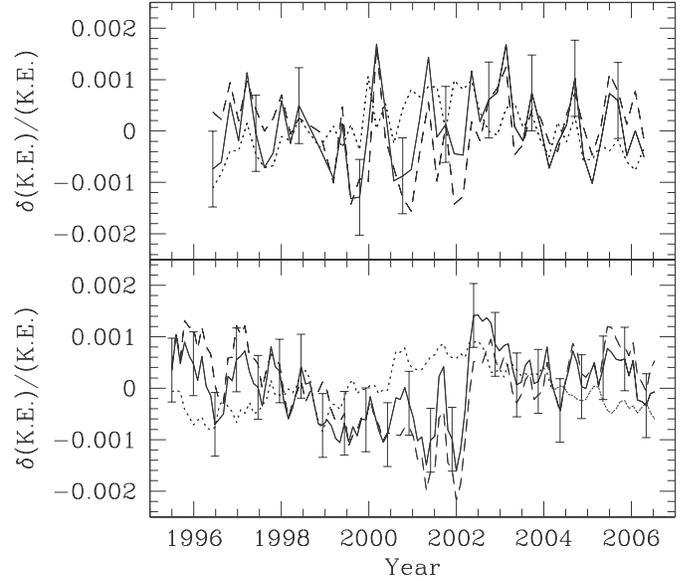} }}
\caption{Temporal variation in rotational kinetic energy
throughout the convection zone.  The continuous lines join values of the 
total kinetic energy  (in units of the temporal mean) in the convection zone 
at different epochs; the 
dashed and dotted lines respectively join corresponding contributions 
from the low ($\le \pi/4$) and high ($\ge\pi/4$) latitude regions.  
For clarity, only a few representative error bars are shown. 
The values in the upper panel were obtained from the MDI data, the 
values in the lower panel from the GONG data.}
\end{figure}

\begin{figure}[t]
\centerline{\resizebox{\figwidth}{!}{\includegraphics{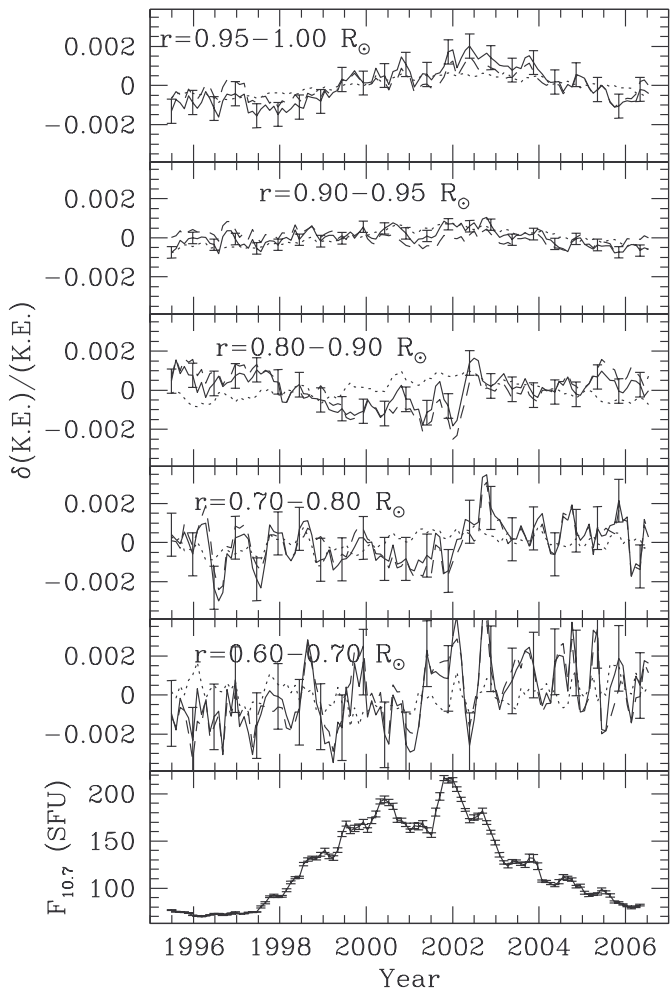} }}
\caption{Relative variation with time of the rotational kinetic energy
in different regions (indicated in each panel) of the Sun, obtained from the 
GONG data. The continuous lines 
indicate the variation over entire spherical shells, and the
dashed and dotted lines respectively indicate the variation in the 
low- ($\le \pi/4$) and high-latitude ($\ge\pi/4$) regions. 
For clarity, only a few representative error bars are shown. The lowest panel
shows the variation in solar activity as measured by the
10.7 cm radio flux, over the same time interval.}
\end{figure}

We show in Figure~1 the temporal variation of the rotational kinetic energy 
of the entire convection zone, together with the contributions from the 
high- and low-latitude regions.  The upper panel
was obtained from the MDI data, the lower panel from GONG.  An oscillatory 
variation with about an 11-year period is evident, although there is some 
difference between the values inferred from the MDI and the GONG data.

Figure~2 depicts the temporal variation in rotational kinetic energy
in different regions of the Sun obtained using the GONG data.
The magnitude of the absolute variation increases with depth, largely
on account of increasing mass density. In order to get a more uniform
comparison amongst different depths, the variation relative to the
mean value of the kinetic energy in each layer is plotted; then the
amplitudes are rather similar.  The contributions from low latitudes
$(\theta > \pi/4)$ and high latitudes $(\theta < \pi/4)$ are also
plotted. The relative magnitude of these variations is also taken with respect
to the mean kinetic energy in the entire latitude range.
In all cases these variations are similar in magnitude.  We note that
the variation in angular velocity is substantially
greater at high latitudes than it is near the equator. 
The effect of that on the variation of the rotational kinetic energy and angular momentum 
is partly compensated by the fact that
the high-latitude regions are closer to the rotation axis, and
contribute less to the moment of inertia, a property which 
is reflected by the factor $\sin^3\theta$ in the
integrands in equations (2) and (3), which on its own integrates to
$5/6 \sqrt{2} \approx 0.589$ and $2/3-5/6 \sqrt{2} = 0.077$ in the low- and
high-latitude regions respectively.

At the bottom of Figure 2 is plotted the 10.7 cm radio flux, which is
an indicator of solar activity.  It is evident that in the outer
layers $(r \ga 0.9 {\rm R}_{\odot})$ of the Sun the kinetic energy is
greatest at epochs of greatest activity, both in the low-latitude and
in the high-latitude regions.  At intermediate depths $(0.7 \la
r/{\rm R}_{\odot} \la 0.9)$ the high-latitude correlation with
activity persists, but at low latitudes the correlation is reversed.
This trend is seen more clearly in Figure~3, in which the correlation 
coefficients are plotted as a function of radius, both for entire 
spherical shells and for the high- and low-latitude regions. 
The amplitude of the variation is the greater in
the low-latitude region, so the behaviour there is an
indicator of the behaviour of the kinetic energy, and the angular
momentum, over the entire spherical shell.  At the greatest depths it
is more difficult to see the trends in Figure 2 because the
uncertainties in the inversions are comparable with the trends
themselves.  However, the covariance of the kinetic energy with the
activity formally changes sign at $r \simeq 0.7{\rm R}_{\odot}$, at
both low and high latitudes, beneath which the low-latitude kinetic
energy variation is again positively correlated with activity and the
high-latitude variation is negatively correlated.  Again, this behaviour 
is visible more clearly in Figure 3. 

These results are consistent with those of Komm et al.~(2003), who
considered the angular-momentum variations over entire spherical
shells.  The relative variations in angular momentum and rotational
kinetic energy, and also, of course, the angular velocity, are of
order 0.1\%.  Given that the total angular momentum and rotational
kinetic energy of the convection zone are $J_{\rm c} \simeq 2.8 
\times 10^{47} {\rm g\, cm}^2 {\rm s}^{-1}$ and $T_{\rm c} \simeq 4.0 
\times 10^{41}$ erg, this
implies that the amplitudes of their variations are about $1 \times 10^{44} 
{\rm~g~cm}^2{\rm s}^{-1}$ and $3 \times 10^{38}~$ erg respectively.

\begin{figure}[t]
\centerline{\resizebox{\figwidth}{!}{\includegraphics{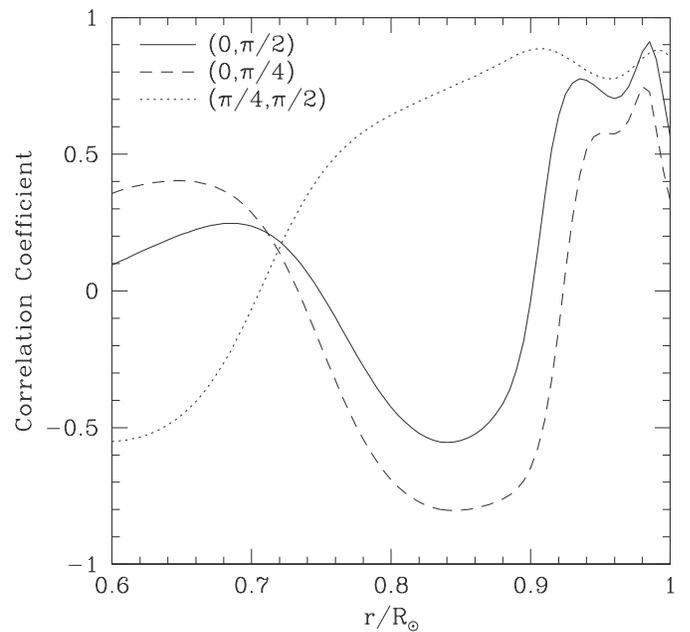} }}
\caption{Correlation coefficients between the temporal
variation in kinetic energy of rotation obtained using GONG data and
the 10.7 cm radio flux, 
plotted as a function of radius. The continuous curve is the coefficient 
for an entire spherical shell; the
dashed and dotted curves are respectively the results for the 
low-latitude ($\le \pi/4$) and high-latitude ($\ge\pi/4$) regions.}
\end{figure}

 The change in sign at $r
\simeq 0.9{\rm R}_{\odot}$ of the correlation of angular momentum with
activity at low latitudes could be associated with the upward
migration of the zonal flow pattern reported by Basu \& Antia (2003):
because the migration timescale from intermediate depths is about half a
cycle period, the flow pattern at intermediate depths is anticorrelated
with that at the surface.  At high latitudes, the angular-momentum
variation is positively correlated with surface activity throughout
the convection zone.  Beneath the convection zone the correlation
appears to be weaker, possibly as a result of the greater
uncertainties in the inferred angular velocity, and may not be
significant.  

\begin{figure}[t]
\centerline{\resizebox{\figwidth}{!}{\includegraphics{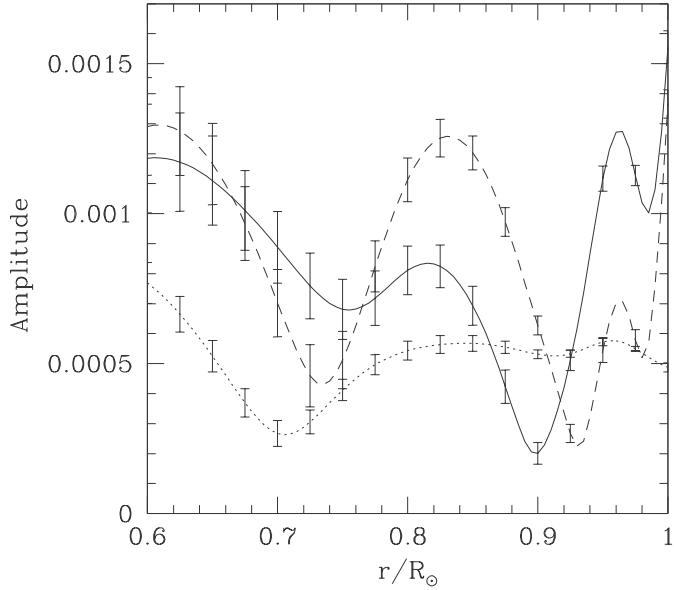} }}
\caption{Amplitude of relative 
variation in rotational kinetic energy, plotted against radius, 
obtained by fitting a sinusoid with a period of 11 years to the inferences 
from GONG data.  The continuous curve  is for the entire
spherical shell, the
dashed and dotted curves respectively are the results for the low ($\le \pi/4$)
and high ($\ge\pi/4$) latitude regions.}
\end{figure}

\begin{figure}[t]
\centerline{\resizebox{\figwidth}{!}{\includegraphics{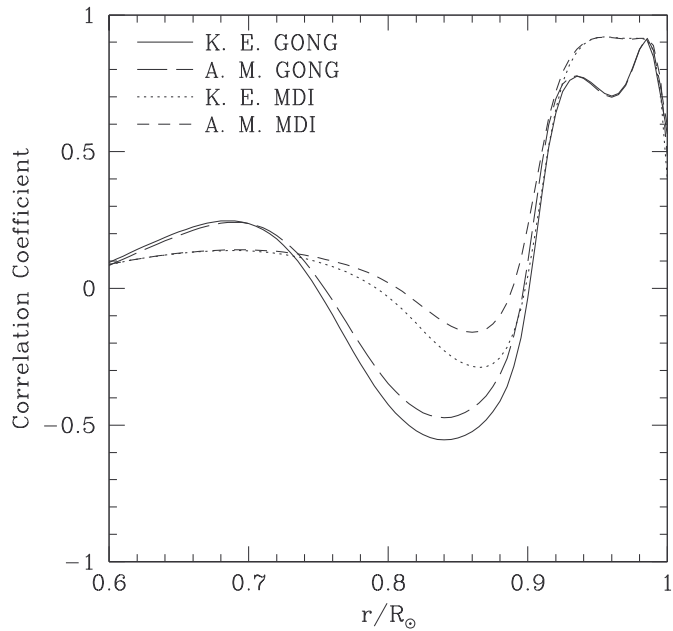} }}
\caption{Correlation coefficient between the the 10.7 cm radio flux and the 
kinetic energy and angular momentum, integrated over
entire spherical shells, plotted against radius. 
The solid and long-dashed curves are
kinetic energy and angular momentum inferred from GONG data,
the dotted and short-dashed curves are from MDI data.}
\end{figure}

To calculate the amplitude of the solar-cycle variation in kinetic energy, we fit
a sinusoid with a period of 11 years to the observed kinetic energy.  Since the
power spectra at most depths show a peak at frequencies corresponding
to approximately this value, we may be justified in looking for such
variation. The resulting amplitude of the relative variation is plotted 
against radius in Figure~4. The dips in the curves occur near the depths 
at which 
the correlation coefficient changes sign (see Figure ~3).
At these depths the amplitude might be expected to 
be small.  Beneath the convection zone the rise in amplitude with depth 
might be merely a reflection of the increasing errors in the inferred 
angular velocity, and is therefore not to be trusted; this is suggested 
partly by the low correlation coefficients evident in Figure 3.

\begin{figure}[t]
\centerline{\resizebox{\figwidth}{!}{\includegraphics{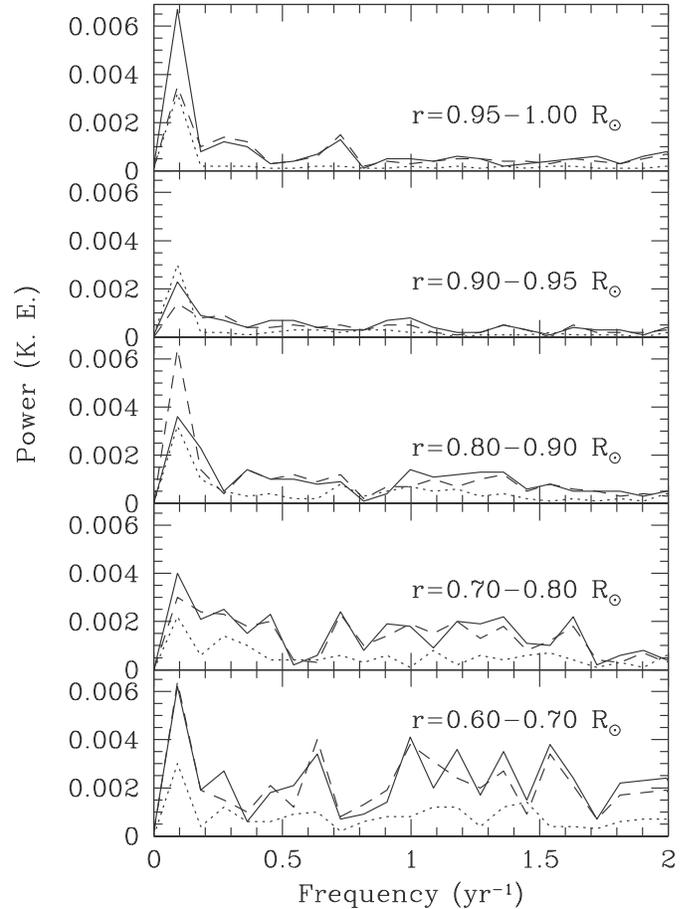} }}
\caption{Power spectrum of temporal
variation in rotational kinetic energy in different depth ranges  (as marked
in each panel).  The solid lines are for the entire
spherical shell, the dashed and dotted lines are for low- ($\le \pi/4$)
and high- ($\ge\pi/4$) latitude regions, respectively.  The seismic data 
were obtained from GONG; the results from MDI are similar, and have been 
omitted for concision.}
\end{figure}

\begin{figure}[t]
\centerline{\resizebox{\figwidth}{!}{\includegraphics{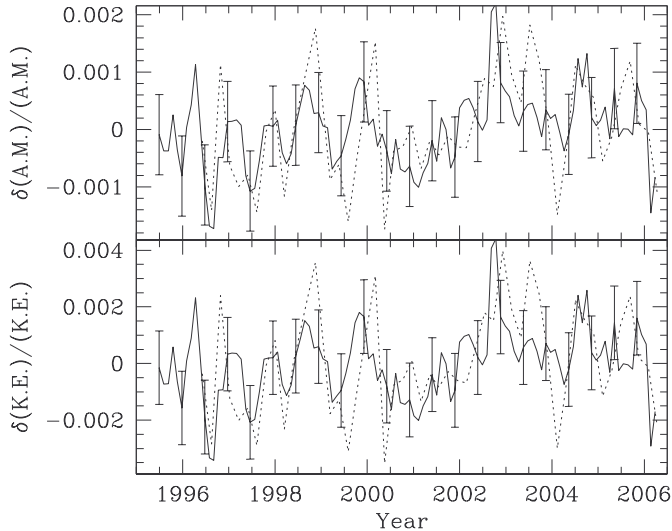} }}
\caption{Temporal variation in rotational kinetic energy
and angular momentum in the low-latitude region with
$0.70R_\odot\le r\le0.74R_\odot$. 
The continuous lines are the result of 
using GONG data, the dotted lines are from using MDI data.
For clarity, only a few representative error bars are shown.}
\end{figure}

We conclude this discussion by once again comparing inferences 
from GONG and MDI.  Figure 5 depicts the correlation coefficients 
between the 10.7 cm radio flux and the  
spherically averaged rotational kinetic energy and angular momentum
obtained from the two data sets. It is evident that there is 
very little difference between the coefficients for 
rotational kinetic energy and angular momentum. There is some 
discrepancy between the inferences from GONG and MDI in the lower part
of the convection zone, a discrepancy similar to one that has been found 
also in other works (e.g., Schou et al.~2002). The reason for that 
discrepancy is not fully understood.  Nevertheless, it does not affect 
the principal conclusions of this paper. 

Howe et al.~(2000b) and Komm et al.~(2003) have reported a 1.3-year 
periodicity in the temporal variation of the angular velocity at low
latitudes in the vicinity of the tachocline.  Here we examine this 
finding further, by seeking such a periodicity in the angular momentum 
or kinetic energy.   We calculate the discrete Fourier transforms 
of the results illustrated in Figure~2 to see whether there is a 
significant peak in the power 
spectra around a frequency of 0.75 yr$^{-1}$.  The outcome is presented 
in Figure 6. Evidently, no such peak is apparent. The power spectra in 
the vicinity of the tachocline appear to contain no particularly 
significant peak, although those in the upper layers of the convection zone 
do contain an obvious peak in the lowest frequency bin allowed in the 
discrete Fourier transform: this corresponds to the solar-cycle variation. 
We have looked at many different depth and latitude ranges; the results are 
all similar to those shown in the figure, irrespective of whether the seismic 
data employed were from GONG or from MDI.  Therefore, apart from a possible 
periodicity characteristic of solar-cycle variation, we find no evidence 
for periodicity at any depth.  For illustrative purposes, we present in 
Figure 7 the temporal variation in the low-latitude region of the 
tachocline ($0.70R_\odot\le r\le 0.74R_\odot$) in
both rotational kinetic energy and angular momentum, inferred from MDI 
as well as GONG data. This is the region in which
Howe et al.~(2000b) and Komm et al.~(2003)
found the greatest amplitude of the 1.3-year oscillations. It is clear from 
the figure that there is reasonable
agreement between GONG and MDI data, and that the variation in angular
momentum is similar to that in the rotational kinetic energy.
However, no significant temporal variation is evident; in particular, 
there is no 
very
clear sign of a 1.3-year oscillation.

\section{Gravitational Multipole Moments due to Internal Rotation}

\begin{table*}
\begin{center}
\caption{Temporal variation in multipole moments}
\begin{tabular}[h]{ccccccc}
\hline
\noalign{\smallskip}
$k$&\multispan2{\hfill $\langle J_k\rangle$\hfill}&
\multispan2{\hfill Phase (years)\hfill}&
\multispan2{\hfill $|\delta J_k|$\hfill}\\
&GONG&MDI&GONG&MDI&GONG&MDI\\
\noalign{\smallskip}
\hline
\noalign{\smallskip}
2&$(2.18\pm0.005)\times10^{-7}$&$(2.22\pm0.009)\times10^{-7}$
&$4.1$&\phantom{1}$8.7$&
$1.0\times10^{-10}$&$9.5\times10^{-11}$\\
4&$-(4.70\pm0.06)\times10^{-9}$&$-(3.97\pm0.09)\times10^{-9}$
&$1.4$&\phantom{1}$0.1$&
$4.9\times10^{-11}$&$7.9\times10^{-11}$\\
6&$-(2.4\pm0.2)\times10^{-10}$&$-(0.8\pm0.3)\times10^{-10}$
&$7.3$&$10.7$&
$4.8\times10^{-12}$&$1.3\times10^{-11}$\\
8&$-(0.8\pm0.6)\times10^{-11}$&$\phantom{-}(1.1\pm0.9)\times10^{-11}$
&$4.5$&\phantom{1}$3.6$&
$1.6\times10^{-11}$&$1.7\times10^{-11}$\\
10&$\phantom{-}(7.1\pm4.0)\times10^{-12}$&$\phantom{-}(7.4\pm4.0)\times10^{-12}$
&$1.6$&\phantom{1}$1.4$&
$7.3\times10^{-12}$&$7.8\times10^{-12}$\\
12&${-}(1.9\pm2.4)\times10^{-12}$&${-}(1.8\pm2.0)\times10^{-12}$
&$6.8$&\phantom{1}$6.9$&
$3.2\times10^{-12}$&$2.6\times10^{-12}$\\
\noalign{\smallskip}
\hline
\end{tabular}
\end{center}
\end{table*}

From our inferences of the angular velocity throughout the solar interior, 
it is possible to calculate the quadrupole and higher-order multipole 
moments of the Sun's gravitational field, and study their possible temporal 
variation. Figure~8 illustrates the even-order moments $J_2$ -- $J_{12}$ 
inferred from both GONG and MDI data.  To 
ascertain 
whether temporal 
variation in these quantities is related to solar activity we have 
calculated the coefficients of their correlation with the 10.7 cm radio 
flux. We 
identify the phase of any possible solar-cycle variation 
by fitting to the moments sinusoids with a period of 11 years.  The results are 
summarized in Table~1, in which are listed the temporal averages of the 
moments as well as the relative amplitudes and phases (in years, relative 
to the phase of the 10.7 cm radio flux) of the fitted sinusoidals.
As one might expect, the quadrupole moment $J_2$ exhibits no noticeable 
temporal variation: the sinusoidal fits have very low (relative) amplitudes, 
and the phases inferred from the GONG and MDI data are different.
There appears to be a substantial difference between the 
mean value of $J_2$  calculated from the GONG and the MDI data. This is most 
probably a result of already known differences in the splitting coefficients 
(Schou et al.~2002).  Similar differences between GONG and MDI estimates
for $J_2$ were found by Pijpers (1998).  We note in passing that 
Emilio et al.~(2007), using measurements by MDI, find strong variation in the 
figure of the Sun, which is contrary to their earlier results 
(Kuhn et al.~1998); however, 
most of the distortion from sphericity is the direct response 
to the centrifugal force on the rotating surface layers, and not from 
the asphericity of the gravitational field.  
The temporal variation in the distortion observed at the solar surface is 
certainly associated with the magnetic field, which itself is correlated with 
the zonal bands of alternate fast and slow rotation, and is probably confined 
to layers near the surface.  Of course, the uncertainty 
in the estimate of the gravitational distortion obtained from the 
shape of the surface of the Sun is much greater than the uncertainty with 
which $J_2$ can be 
obtained 
from seismic measurements.

\begin{figure}[t]
\centerline{\resizebox{\figwidth}{!}{\includegraphics{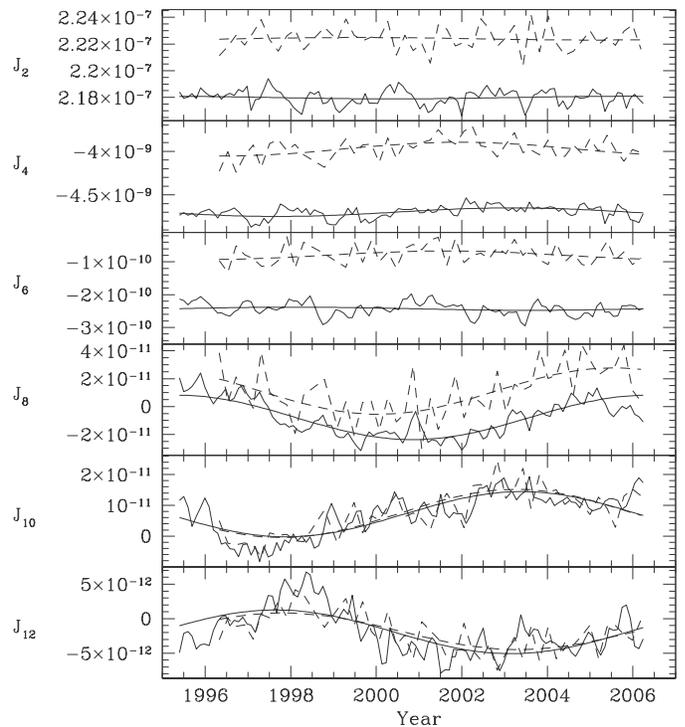} }}
\caption{Temporal variation in (even) multipole moments $J_2$ -- $J_{12}$. 
The solid lines are the results from using GONG data, 
dashed lines from using MDI data. In all cases a sinusoidal fit, with a 
period of 11 years, is also shown.}
\end{figure}

Kuhn et al.~(1998) found significant temporal variation in the
$P_4(\cos\theta)$ component of distortion at the solar surface, although
their errors are much larger in this component. Accordingly, we seek a 
corresponding variation in $J_4$. It appears in Figure~8 that there is indeed 
some systematic variation, for inferences from both GONG and MDI data  
behave similarly.  However, the amplitude of the sinusoid fitted to 
the GONG data is only about 60\% of that of the corresponding MDI sinusoid, 
and, moreover, the phases differ by about 1.3 years. The correlation 
coefficients between the inferred values of $J_4$ and the
10.7 cm radio flux are only $0.29$ and $0.51$ for GONG
and MDI data respectively, although these quite low values appear to be 
a consequence mainly of the phase difference.   
However, the results do suggest that $J_4$ may 
undergo temporal variation with an amplitude of about $10^{-10}$, which is 
2\% of the mean,  and that this variation is correlated with solar activity.
The variation is much less than what Kuhn et al.~(1998) found in
the $P_4(\cos\theta)$ component of the distortion from sphericity of the shape 
of the Sun's surface;  the difference could be in part 
a product of large errors in the distortion observed, but it is likely to be 
due mainly to the direct effect of the centrifugal force, and also the magnetic 
field and accompanying velocity and temperature variations associated with 
solar activity in the low-density surface layers, which influence the figure 
but hardly modify the gravitational field, even though the kernels for 
the high-order multipole moments are more concentrated near the 
surface.   The mean value of $J_4$ in 
Table~1 is close to that found by Roxburgh (2001) and Mecheri et al.~(2004), who  
used 
rough 
approximations to the seismically inferred rotation rate.  

The variations in $J_6$ that we have inferred from the GONG and MDI data do not 
agree: whereas the MDI results suggest a clear variation which is
correlated with solar activity, the GONG results reveal none. The mean values 
of $J_6$ differ by a factor 3, although they are at least of the same order 
as that found by Roxburgh (2001).  

The variation in $J_8$ is anticorrelated with 
solar activity, and now the variations inferred from both GONG and MDI data
look similar. However, the mean values do not even have the same sign. 
With $J_{10}$ and $J_{12}$ there is good agreement between GONG and MDI.
The magnitudes of $J_8$ and higher-order moments are comparable with their 
temporal variations, although we hasten to add that they are comparable 
also with the estimated 1-$\sigma$ errors in the individual estimates. 
Therefore they may not be significant.  But we have presented them because 
they do tend to show definite temporal trends which are similar in both the 
GONG and MDI inferences.  Higher-order multipole moments also exhibit 
variation, but we do not present them because their reliability is 
certainly questionable.

We point out that the contribution from rotation to each multipole moment 
can be expressed as a bilinear function of the odd degeneracy splitting 
coefficients of the seismic modes, a function which depends also on the 
density $\rho(r)$.  Therefore, because the density of the same reference 
solar model was used to compute all the moments, whether the seismic data used 
were from GONG or MDI, all the differences amongst the inferences must 
have their origin solely in the differences amongst the measured splitting 
coefficients.

\section{Summary and Conclusions}

We have studied the temporal variation in the rotational kinetic energy and
angular momentum of the Sun in different regions of the convection
zone. We find that in
the outer layers ($r\ga 0.9R_\odot$) the variation is well correlated
with solar activity. At high latitudes this correlation continues
throughout the convection zone, while at low latitudes (and also when
integrated over all latitudes), there is a transition to
anticorrelation around $r=0.9R_\odot$. Through the bulk of the
convection zone the high latitudes seem to be rotating faster during
the solar activity maximum, and the equatorial regions rotate 
more slowly.  It is difficult to draw any definite
conclusions about the temporal variation below the convection zone, 
however, because the errors are large.

\begin{table}
\begin{center}
\caption{Rotational kinetic energy, thermal energy and the estimated
rms magnetic field using Eq.~(5) in different layers}
\begin{tabular}[h]{@{}ccrc}
\hline
\noalign{\smallskip}
Radial extent&$\!\!\!$Mean rot.~KE&rms~$ B$\hfill&Thermal energy\\
&(erg)&(G)&(erg)\\
\noalign{\smallskip}
\hline
\noalign{\smallskip}
$0.99<r/R_\odot<1.00$&$5.3\times10^{37}$&250&$6.2\times10^{40}$\\
$0.95<r/R_\odot<0.99$&$6.5\times10^{39}$&1400&$2.4\times10^{43}$\\
$0.85<r/R_\odot<0.95$&$9.3\times10^{40}$&3700&$1.4\times10^{45}$\\
$0.75<r/R_\odot<0.85$&$2.0\times10^{41}$&6100&$7.6\times10^{45}$\\
$0.65<r/R_\odot<0.75$&$2.6\times10^{41}$&7900&$2.2\times10^{46}$\\
\noalign{\smallskip}
\hline
\end{tabular}
\end{center}
\end{table}

The amplitude of the temporal variation in the rotational kinetic energy
integrated over the entire convection zone is of order of $3\times10^{38}$
erg.   This implies a rate of variation whose amplitude is about 
$5\times10^{30}$ erg s$^{-1}$. 
It is not clear if this variation is compensated for by opposite
variation in deeper layers or whether it represents exchange of
energy between rotation and magnetic field or some other form.
It is interesting to note that this value is roughly comparable to 
(actually about 3 times) the amplitude of the observed variation in the 
solar irradiance, which is about 0.04\%, although the correspondence might 
well be fortuitous.  It is, however, tempting to speculate that
the temporal variation in rotational kinetic energy, and the magnetic energy, 
has some role to play in the irradiance variation. 
A part of the variation in the rotational kinetic energy would
be expected to be compensated by the magnetic energy stored
in the solar convection zone. If that were the case at all radii, we can 
estimate the spatial variation of the rms magnetic field using the relation
\begin{equation}
{B^2\over 8\pi}4\pi r^2\delta r\approx \delta T.
\end{equation}
The total variation $\delta T$ in $T$ over the solar cycle is about 
0.002 $T$, where $T(r_1,r_2)$ is the mean rotational kinetic energy in the 
region between radii $r_1$ and $r_2$.  Using this, together with the 
mean kinetic energy integrated over all latitudes within several spherical 
shells, we list in Table~2 these admittedly very crude estimates of the 
magnetic field. 

The gravitational quadrupole moment $J_2$ does not exhibit any significant temporal
variation, whereas some other, higher-order, multipole moments do. 
The variation in $J_4$ is correlated with solar activity, but the 
amplitudes of variation seen in GONG and MDI data are somewhat different. 
Clear variation in $J_8$ -- $J_{12}$ is inferred from both GONG and MDI 
data, the variation in $J_{10}$ being correlated with solar activity and that 
in $J_8$ and $J_{12}$ anticorrelated.
The absence of temporal variation in $J_2$ is
consistent with observed absence of variation in the
$P_2(\cos\theta)$ component of distortion at the solar surface
(Kuhn et al.~1998).  This is to be expected, since $J_2$ is determined  
essentially by the spherically averaged component of the angular velocity, 
which suffers little relative variation.
The higher-order multipole moments are determined also by the latitudinally
varying components of the angular velocity, which are known to suffer 
temporal variation correlated with solar activity.
Note that the predominant contribution to $J_2$ arises from regions near the
base of the convection zone (Gough, 1981; Pijpers 1998), whereas the 
higher-order multipole moments arise largely from contributions from 
outer layers of the Sun, where the temporal variation in the rotation rate 
is not insubstantial.  Of course, the magnitude of $J_2$ is much larger than 
that of the higher multipole moments, and it is likely that $J_2$ also has 
temporal variations with comparable absolute magnitude. 

\begin{acknowledgements}
This work  utilized data obtained by the Global Oscillation
Network Group (GONG) project, managed by the National Solar Observatory,
which is
operated by AURA, Inc. under a cooperative agreement with the
National Science Foundation. The data were acquired by instruments
operated by the Big Bear Solar Observatory, High Altitude Observatory,
Learmonth Solar Observatory, Udaipur Solar Observatory, Instituto de
Astrofisico de Canarias, and Cerro Tololo Inter-American Observatory.
This work also utilises data from the Solar Oscillations
Investigation/ Michelson Doppler Imager (SOI/MDI) on the Solar
and Heliospheric Observatory (SOHO).  SOHO is a project of
international cooperation between ESA and NASA.
We are grateful to J. Christensen-Dalsgaard, T. Duvall, L. Gizon
and P. Scherrer for helpful discussions and valuable comments.
SMC thanks DAE-BRNS for support under the Senior
Scientist Scheme and is also grateful to the Institute of Astronomy,
Cambridge for supporting his visits during the course of this study.

\end{acknowledgements}


\begin{thebibliography}{}

\bibitem[]{ab00}
Antia, H. M. \& Basu, S. 2000, ApJ, 541, 442

\bibitem[]{ab01}
Antia, H. M. \& Basu, S. 2001, ApJ, 559, L67

\bibitem[]{abc98}
Antia, H. M., Basu, S. \& Chitre, S. M. 1998, MNRAS, 298, 543


\bibitem[]{act00}
Antia, H. M., Chitre, S. M. \& Thompson, M. J. 2000, {A\&A}, {360}, 335

\bibitem[]{ba03}
Basu, S. \& Antia, H. M. 2003, ApJ, 585, 553

\bibitem[]{ba06}
Basu, S. \& Antia, H. M. 2006, in proc.\ SOHO 18/GONG 2006/HELAS I
workshop on Beyond the Spherical Sun, Eds.\ K. Fletcher and M. J.
Thompson, ESA SP-624, p.~128






\bibitem[]{emi07}
Emilio, M., Bush, R. I., Kuhn, J. \& Scherrer, P. 2007, ApJ, 560, 161


\bibitem[1981]{dog81}
Gough, D. O. 1981, {MNRAS,} {196}, 731

\bibitem[1982]{dog82}
Gough, D. O. 1982, Nature, 298, 334 

\bibitem[1982]{dog07}
Gough, D. O. 2007, in Dynamics of the Solar Tachocline, ed. D.W. Hughes, 
R. Rosner \& N.O. Weiss, Cambridge University Press

\bibitem[]{hil96}
Hill, F., et al.~1996, Science, 272, 1292

\bibitem[]{how80}
Howard, R. \& LaBonte, B. J. 1980, ApJ, 239, L33

\bibitem[]{how00a}
Howe, R., \jcd, J., Hill, F., Komm, R. W.,
Larsen, R. M., Schou, J., Thompson, M. J. \& Toomre, J. 2000a, ApJ,
533, L163

\bibitem[]{how00b}
Howe, R., \jcd, J., Hill, F., Komm, R. W.,
Larsen, R. M., Schou, J., Thompson, M. J. \& Toomre, J. 2000b, Sci,
287, 2456

\bibitem[]{how05}
Howe, R., \jcd, J., Hill, F., Komm, R. W.,
Schou, J., \& Thompson, M. J. 2005, ApJ 634, 1405


\bibitem[]{hcdhkstt07}
Howe, R., Christensen-Dalsgaard, J., Hill, F., Komm, R., Schou, J., 
Thompson, M. J.,  Toomre, J.   2007, Adv. Space Res., 40, 915

\bibitem[]{how06}
Howe, R., Rempel, M., \jcd, J., Hill, F., Komm, R. W.,
Larsen, R. M., Schou, J., \& Thompson, M. J. 2006, ApJ 649, 1155

\bibitem[]{kom03}
Komm, R., Howe, R., Durney, B. R. \& Hill, F. 2003, ApJ, 586, 650

\bibitem[1998]{kuhn98} Kuhn, J. R., Bush, R. I., Scheick, X. \& Scherrer, P.
1998, Nature, 392, 155

\bibitem[2004]{mac04}
Mecheri, R., Abdelatif, T., Irbah, A., Provost, J. \& Berthomieu, G.
2004, Solar Phys., 222, 191

\bibitem[]{pij98}
Pijpers, F. P. 1998, {MNRAS,} {297}, L76

\bibitem[2001]{rox01}
Roxburgh, I. W. 2001, A\&A, 377, 688

\bibitem[]{sch99}
Schou, J. 1999, ApJ, 523, L181

\bibitem[]{sch98}
Schou, J. et al.~1998, ApJ, 505, 390

\bibitem[]{sch02}
Schou, J., Howe, R., Basu, S., \jcd, J., Corbard, T., Hill, F., Komm, R.,
Larsen, R. M., Rabello-Soares, M. C. \& Thompson, M. J.  2002, ApJ, 567, 1234

\bibitem[]{schw47}
Schwarzschild, M.~1947, ApJ, 106, 427

\bibitem[]{ps50}
Sweet, P. A.  ,1950,  {MNRAS,}  110, 548

\bibitem[]{tho96}
Thompson, M. J. et al.~1996, Science, 272, 1300

\bibitem[1981]{ulr81}
Ulrich, R. K. \& Hawkins, G. W. 1981, ApJ, 246, 985 (Erratum ApJ 249, 831) 

\bibitem[]{vor02}
Vorontsov, S. V., \jcd, J., Schou, J., Strakhov, V. N.
\& Thompson, M. J. 2002, Science, 296, 101

\end{thebibliography}
\end{document}